\documentclass[12pt]{article}
\def\1{\'{\i}}

\usepackage{amsmath}
\usepackage{amsbsy}
\usepackage{fullpage}

\title{On Relativistic Multipole Moments of Stationary Spacetimes}
\author{Francisco Frutos-Alfaro \\ 
\small{School of Physics and Space Research Center of the University of 
Costa Rica} \\ Michael Soffel \\
\small{Technical University Dresden and Lohrmann Observatory}} 
\date{\today}

\begin{document}
\maketitle

\begin{abstract} 
Among the known exact solutions of Einstein's vacuum field equations the 
Manko-Novikov and the Quevedo-Mashhoon metrics might be suitable ones for the 
description of the exterior gravitational field of some real non-collapsed 
body. A new proposal to represent such exterior field is the stationary 
$ q $-metric. In this contribution, we computed by means of the 
Fodor-Hoenselaers-Perj\'es formalism the lowest ten relativistic multipole 
moments of these metrics. Corresponding moments were derived for the static 
vacuum solutions of Gutsunayev-Manko and Hern\'andez-Mart{\'{\i}}n. 
A direct comparison between the multipole moments of these non-iso\-metric 
spacetimes is given. 
\end{abstract}

%\newpage

\section{Introduction}

\noindent
In 1968, Ernst developed a complex procedure to simplify the Einstein field 
equations (EFE) for a stationary Weyl-Lewis-Papapetrou (WLP) type metric 
by means of two complex potentials \cite{Ernst1,Ernst2}. After this seminal 
work, several methods to find new solutions of the EFE were developed by 
Hoenselaers, Kinnersley and Xanthopoulos (HKX) \cite{HKX}, Belinsky and Zakharov
\cite{BZ1,BZ2}, among others. The HKX method was employed by Quevedo and
Mashhoon to find new stationary solutions from the Erez-Rosen spacetime as 
seed metric \cite{QM1,QM2,Quevedo3}. The Erez-Rosen metric is an exact vacuum 
solution of EFE representing a static metric with a mass and a quadrupole 
parameter ($ M $ and $ Q $) \cite{Carmeli}. The Quevedo-Mashhoon 
(QM) metric is an axially-symmetric stationary vacuum solution with parameters 
$ M, \, a $ and $ q_n $ with $ n $ being an (even) integer 
($ q_2 $ can be identified with a quadrupole parameter) \cite{Boshkayev}. 
The Manko-Novikov (MN) metric was derived by solving the Ernst equations using 
the techniques developed by Yamazaki, Cosgrove, Dietz and Hoenselaers 
\cite{Yamazaki2,Cosgrove,Dietz}. This metric has parameters $ k, \, \alpha $ 
(representing mass and rotation parameter, respectively) and $ \alpha_n $ 
($ \alpha_2 $ representing the quadrupole) \cite{Manko}. Quevedo is proposing 
the stationary version of the $ q $-metric to describe the gravitational field 
of compact stars \cite{Toktarbay,Quevedo5,Quevedo6}. 
Among the exact stationary vacuum solutions of EFE these metrics could be 
appropriate ones for the description of the exterior gravitational field of 
some real non-collapsed body simply because of their large freedom in choosing 
the multipole moments of the exterior gravitational field. Note, however, 
that the spin moments of higher order cannot be chosen freely but are totally 
determined by the set of parameters just mentioned; they might be far from 
those of a realistic body ({e.g.}, Panhans and Soffel, and Teyssandier 
\cite{Panhans,Teyssandier}). 

\noindent
Geroch and Hansen (GH) \cite{Geroch,Hansen} have provided a definition for mass 
and spin multipole moments for a stationary spacetime with asymptotic flatness.
Thorne, Simon and Beig gave definitions of relativistic multipole moments 
\cite{Simon,Thorne} even for non-stationary spacetimes. G\"ursel proved that 
the GH multipole moments are equivalent to the Thorne moments for stationary 
systems \cite{Guersel}. From the Ernst formalism, Fodor, Hoenselaers and 
Perj\'es (FHP) found an elegant method to derive explicit expressions for the 
multipole moments of a given stationary (axially symmetric) spacetime with 
asymptotic flatness \cite{Fodor}. Later, this method was generalized by 
Hoenselaers and Perj\'es \cite{Hoenselaers}. Another method for deriving the 
relativistic multipole moments was devised by Ryan \cite{Ryan}.

\noindent
In this paper, the MN and the QM metrics are briefly reviewed in Section 2. 
The stionary version of the q-metric is introduced in this section too.
We then derive the lowest relativistic multipole moments by means 
of the FHP-method (in our notation $ {\cal M}_{2 l} $ and $ {\cal S}_{2 l + 1} $ 
denote the GH mass and spin moments for the axially symmetric case) in 
section 3. As limits the corresponding moments of the Schwarzschild, Kerr and 
Erez-Rosen metrics are contained. In Section 4, for three static cases, 
the Gutsunayev-Manko and Hern\'andez-Mart{\'{\i}}n (I and II) metrics, 
corresponding mass moments were rederived and compared with those from 
the MN, QM and q-metrics. We wrote a REDUCE program to get the first eleven 
multipole moments. Some conclusions are given in the last section. 

%\newpage

\section{The Stationary Metrics}

\noindent
The stationary spacetime is represented by the WLP metric in 
prolate spheroidal coordinates $ (t, \, x, \, y, \phi) $ 

\begin{eqnarray}
\label{stationary}
ds^2 & = & - f (d t - \omega d \phi)^2 \\
& + & \frac{\sigma^2}{f}
\left[{\rm e}^{2 \gamma} (x^2 - y^2) \left(\frac{d x^2}{x^2 - 1}
+ \frac{d y^2}{1 - y^2} \right) + (x^2 - 1)(1 - y^2) d \phi^2 \right] ,
\nonumber
\end{eqnarray}

\noindent
where

\begin{eqnarray}
\label{sigma}
\sigma^2 & = & {m^{2} - a^{2}} . \nonumber
\end{eqnarray}

\noindent
The functions $ f $ and $ \omega $ are related with the twist scalar $ \Omega $ 
through

\begin{equation}
\label{twist}
f^2 \nabla \omega = \rho {\boldsymbol{\phi}} \times \nabla \Omega ,
\end{equation}

\noindent
where $ {\boldsymbol{\phi}} $ as the azimuthal unit vector and $ \rho $ the 
cylindrical coordinate. 

\noindent
The Ernst potential is $ {\cal E} = f + i \Omega $, and the Ernst 
function is given by

\begin{eqnarray}
\label{qm0}
\xi & = & \frac{1 + {\cal E}}{1 - {\cal E}}
\end{eqnarray}

\noindent
This Ernst function and its inverse $ \xi^{-1} $ fulfill the Ernst equation

\begin{equation}
\label{Ernst}
({\xi} {\xi}^{\star} - 1) \nabla^2 {\xi} = 2 {\xi}^{\star} [\nabla {\xi}]^2 .
\end{equation}

\noindent
There are several techniques to solve this Ernst equation, see for instance 
\cite{HKX,BZ1,BZ2,Yamazaki2,Cosgrove,Dietz}.   

%\newpage

\subsection{The Manko-Novikov Metric}

\noindent
This metric represents the spacetime of a rotating massive object with mass and
spin multipole moments. This metric has parameters: $ k, \, \alpha $, and 
$ \alpha_{n}, \, n > 1 $. The parameters $ k $ and $ \alpha $ are related 
with the mass and the rotation, respectively. 
It contains the following exterior metrics

\begin{itemize}
\item Schwarzschild, $ \alpha = \alpha_{n} = 0 $,
\item Kerr, $ \alpha_{n} = 0 $.
\end{itemize}

\noindent
To compare it with the Kerr or QM metric one has to set

\begin{eqnarray}
\label{defmn}
M & = & k \left(\frac{1 + \alpha^2}{1 - \alpha^2} \right) \nonumber \\
a & = & - \frac{2 k \alpha}{1 - \alpha^2} \\
Q & = & - k^3 \alpha_2 \nonumber \\
k^2 & = & \sigma^2 \nonumber
\end{eqnarray}

\noindent
If there is no rotation, $ \alpha = 0 $, then $ k $ represents the mass. 
This solution has an infinite set of relativistic mass and spin multipoles. 
They correspond to the Kerr ones if $ \alpha_n = 0 $.

\noindent

\noindent
The Ernst function is

\begin{eqnarray}
\label{mn1}
\xi & = & {\rm e}^{2 \psi} \frac{{\cal A}_{-}}{{\cal A}_{+}} 
\end{eqnarray}

\noindent
where

\begin{eqnarray}
\label{mn2}
{{\cal A}_{\mp}} & = & x (1 + \eta_1 \eta_2) 
+ i y (\eta_2 - \eta_1) \mp (1 - i \eta_1)(1 - i \eta_2) .
\end{eqnarray}

\noindent
From this Ernst function (\ref{mn1}), employing methods developed by 
Yamazaki, Cosgrove, Dietz and Hoenselaers \cite{Yamazaki2,Cosgrove,Dietz}, 
they generated the following metric potentials

\begin{eqnarray}
\label{MankoN}
f & = & \frac{A}{B} {\rm e}^{2 \psi} , \nonumber \\
\omega & = & 
2 k \frac{C}{A} {\rm e}^{- 2 \psi} - 4 k \frac{\alpha}{1 - \alpha^2} ,  \\
{\rm e}^{2 \gamma} & = & 
\frac{A {\rm e}^{2 {\chi}}}{(1 - \alpha^2)^2 (x^2 - 1)} . \nonumber 
\end{eqnarray}

\noindent
where

\begin{eqnarray}
\label{mn3}
A & = & (x^2 - 1)(1 + \eta_1 \eta_2)^2 
- (1 - y^2)(\eta_2  - \eta_1)^2 , \nonumber \\
B & = & [x + 1 + (x - 1) \eta_1 \eta_2]^2 
+ [(1 + y) \eta_1 + (1 - y) \eta_2]^2 , \nonumber \\
C & = & (x^2 - 1)(1 + \eta_1 \eta_2) 
[\eta_2 - \eta_1 - y (\eta_1 + \eta_2)] \\
& + & (1 - y^2)(\eta_2 - \eta_1) [1 + \eta_1 \eta_2 + x (1 - \eta_1 \eta_2)] , 
\nonumber \\
\psi & = & \sum^{\infty}_{n = 1} \alpha_n \frac{P_n}{R^{n + 1}} . \nonumber
\end{eqnarray}

\noindent
The functions $ \eta_1 , \, \eta_2 , \, \chi_1 $ and $ \chi_2 $ are 

\begin{eqnarray}
\label{mn4}
\eta_1 & = & - \alpha {\rm e}^{- 2 \chi_1} , \nonumber \\
\eta_2 & = & \alpha {\rm e}^{2 \chi_2} , \\
\chi_1 & = & \sum^{\infty}_{m = 1} \sum^{m}_{n = 0} \alpha_m
\left[(x - y) \frac{P_n}{R^{n + 1}} - 1 \right] , \nonumber \\
\chi_2 & = & \sum^{\infty}_{m = 1} \sum^{m}_{n = 0} \alpha_m
\left[(- 1)^{m - n + 1} (x + y) \frac{P_n}{R^{n + 1}} + (- 1)^n \right] , \nonumber 
\end{eqnarray}

\noindent
with following definitions

\begin{eqnarray}
\label{mn5}
R & = & \sqrt{x^2 + y^2 - 1} , \\
P_{n} & = & P_{n} \left(\frac{x y}{R} \right) . \nonumber  
\end{eqnarray}

\noindent
The $ \gamma $ function is not involved in the calculation of the relativistic 
multipole moments. The interested reader may consult the references. 
In section 3, the Ernst function $ \xi $ will be employed to determine 
the mass and spin multipole moments of the MN spacetime.

%\newpage

\subsection{The Quevedo-Mashhoon Metric}

\noindent
The QM metric is a stationary axisymmetric solution of the vacuum 
EFE. This spacetime has parameters: $ M, \, a $, and $ q_{n}, \, n > 1 $, which 
represent the mass, the rotation parameter and additional multipole moment 
parameters of the object. It has another parameter, the Zipoy-Voorhees 
parameter, which we set $ \delta = 1 $. In general, it presents a naked 
singularity. It contains the following exterior metrics

\begin{itemize}
\item Schwarzschild, $ a = q_{n} = 0 $,
\item Erez-Rosen (ER), $ a = 0 , \, q = q_2 , \, q_{n} = 0, \, (n > 2) $ 
\cite{Carmeli},
\item Kerr, $ q_{n} = 0 $,
\item Hartle-Thorne (HT), $ a $ and  $ q $ small, and $ q_{n} = 0, \, (n > 2) $ 
\cite{Boshkayev}.
\end{itemize}

\noindent
In our notation the parameters $ m \equiv G M / c^2 $ and $ a $ have the 
dimension of a length and the parameters $ q_{n} $ are dimensionless.

\noindent
This solution is characterized by an infinite set of relativistic mass
and spin multipoles. They correspond to the Kerr ones if $ q_{n} = 0 $, and to
the Erez-Rosen ones if $ a = 0 $ \cite{Quevedo3}. Boshkayev {\it et al.} have 
shown how to get the HT from the QM metric \cite{Boshkayev}. The HT metric 
is an approximate solution of the EFE for an object with three parameters, 
mass, spin and quadrupole moment. Although this solution is intended for slow 
rotation, it also has been employed to model fast rotating neutron stars. 
It has been argued ({e.g.} Boshkayev \cite{Boshkayev}) that the QM spacetime 
could be applied to model real astrophysical objects, such as neutron stars. 

\noindent
From two HKX transformations on the static Ernst potential 
$ {\cal E}_0 = {\rm e}^{- 2 \psi} $ for the ER spacetime as seed metric, Quevedo 
and Mashhoon found the new Ernst potential, which is given by 
\cite{QM1,Quevedo3}

\begin{eqnarray}
\label{qm1}
\xi & = & \frac{(a_{+} + i b_{+}) + (a_{-} + i b_{-}) {\rm e}^{- 2 \psi}}
{(a_{+} + i b_{+}) - (a_{-} + i b_{-}) {\rm e}^{- 2 \psi}} , 
\end{eqnarray}

\noindent
From the Ernst function (\ref{qm1}), they found the following functions 
\cite{Boshkayev,QM1,QM2,Quevedo3}

\begin{eqnarray}
\label{qm3}
f & = & \frac{\cal A}{\cal B} {\rm e}^{- 2 \psi} , \nonumber \\
\omega & = & - 2 \left(a
+ \sigma \frac{\cal C}{\cal A} {\rm e}^{2 \psi} \right) ,  \\
{\rm e}^{2 \gamma} & = & \frac{1}{4} \left(1 + \frac{m}{\sigma} \right)^{2}
\frac{\cal A}{(x^{2} - 1)} {\rm e}^{2 {\chi}} , \nonumber
\end{eqnarray}

\noindent
The functions $ {\cal A}, \, {\cal B} $ and $ {\cal C} $ are given by

\begin{eqnarray}
\label{qm4}
{\cal A} & = & a_{+} a_{-} + b_{+} b_{-} , \nonumber \\
{\cal B} & = & a_{+}^{2} + b_{+}^{2} , \\
{\cal C} & = & \left[ x (1 - y^{2})(\lambda + \eta) a_{+}
+ y (x^{2} - 1)(1 - \lambda \eta) b_{+} \right] , \nonumber
\end{eqnarray}

\noindent
where

\begin{eqnarray}
\label{qm5}
a_{\pm} & = & x (1 - \lambda \eta) \pm (1 + \lambda \eta) , \nonumber \\
b_{\pm} & = & y (\lambda + \eta) \mp (\lambda - \eta) ,
\end{eqnarray}

\noindent
with

\begin{eqnarray}
\label{qm6}
\lambda & = & \alpha {\rm e}^{2 \delta_{+}} , \nonumber \\
\eta & = & \alpha {\rm e}^{2 \delta_{-}} , \nonumber \\
\alpha a & = & {\sigma - m} .
\end{eqnarray}

% \newpage

\noindent
For the QM metric the potentials $ \psi $ and $ \delta_{\pm} $ are 
of the form \cite{Quevedo1,QM3} 

\begin{eqnarray}
\label{psidelta}
\psi & = & \sum^{\infty}_{n = 1} (-1)^{n} q_{n} P_{n}(y) Q_{n}(x) , \\
\delta_{\pm} & = & \sum^{\infty}_{n = 1} (- 1)^{n} q_n \left[ (\pm 1)^n 
\left( \frac{1}{2} \ln{\left[\frac{(x \pm y)^{2}}{x^{2} - 1} \right]}
- Q_1(x) \right)+ P_{n}(y) Q_{n - 1}(x) \right. \\
& + & \sum^{n - 1}_{k = 1} \left. (\pm 1)^{k} P_{n - k}(y)
(Q_{n - k + 1}(x) - Q_{n - k - 1}(x)) 
{\phantom{\frac{(1)^n}{2}} \!\!\!\!\!\!\!\!\!\!\!\!} \right] . \nonumber
\end{eqnarray}

\noindent
The functions $ P_{l}(y) $ and $ Q_{l}(x) $ are Legendre polynomials of the 
first and second kind, respectively. The general form of the potential $ \chi $ 
was determined by Quevedo \cite{Quevedo2,Quevedo3}. To find the relativistic 
multipole moments, one does not need this $ \chi $ function. The interested 
reader may consult the references.

\noindent
With the Computer Algebra System REDUCE \cite{Hearn}, we were able to check 
the validity of the Ernst equation directly without employing 
HKX-transformations. In the next section, the Ernst function $ \xi $ will be 
employed to determine the relativistic multipole moments of the QM spacetime.

\newpage

\subsection{The Stationary $ q $-Metric}

\noindent
The $ q $-metric is a generalization of the Schwarzschild metric with 
quadrupole parameter. The static version in spherical coordinates is given 
by \cite{Quevedo5,FQP}

\begin{eqnarray}
\label{qmetric}
d {s}^2 & = & - h^{1 + q} d t^2
+ h^{- q} 
\left[\left(1 + \frac{m^2 \sin^2{\theta}}{r^2 h} \right)^{- q (2 + q)} 
\left( \frac{d r^2}{h} + r^2 d {\theta}^2 \right) 
+ r^2 \sin^2{\theta} d \phi^2 \right] ,
\end{eqnarray}

\noindent
where $ h = 1 - {2 m}/{r} $. It was obtained from the Zipoy-Voorhees (ZV) 
transformation with $ \delta = 1 + q $. From the parameters $ m $ and $ q $, 
the mass and the quadrupole moment of the object are given by 
$ M_0 = (1 + q) m $, $ M_2 = - m^3 q (1 + q) (2 + q)/3 $, respectively. 
It is the simplest static metric with mass and quadrupole parameters. 
The rotating version of this metric has the following Ernst potential in 
prolate spheroidal coordinates

\begin{eqnarray}
\label{qmernst}
{\cal E} & = & \left(\frac{x - 1}{x + 1} \right)^{q}
\left[\frac{x - 1 + (x^2 - 1)^{- q} d_{+}}{x + 1 + (x^2 - 1)^{- q} d_{-}} \right] , 
\end{eqnarray}

\noindent
where 

\begin{eqnarray}
\label{qmernst2}
d_{\pm} & = & - \alpha^2 (x \pm 1) h_{+} h_{-} (x^2 - 1)^{- q}
+ i \alpha [y (h_{+} + h_{-}) \pm (h_{+} - h_{-})] , \nonumber \\
h_{\pm} & = & (x \pm y)^{2 q} . \nonumber
\end{eqnarray}

\noindent
The parameter $ \alpha $ is related with the rotation parameter $ a $ 
by (\ref{qm6}). The prolate spheroidal coordinates are linked to the spherical 
coordinates through

\begin{eqnarray}
\label{prolco}
{\sigma} x & = & {r} - {m} \\
y & = & \cos{\theta} . \nonumber
\end{eqnarray}

\noindent
The rotating metric can be read off from the general QM metric with 
ZV parameter. The Papapetrou potentials are 

\begin{eqnarray}
\label{q0}
f & = & \frac{{\cal A}}{{\cal B}} , \nonumber \\
\omega & = & - 2 \left(a + \sigma \frac{{\cal C}}{{\cal A}} \right),  \\
{\rm e}^{2 \gamma} & = & \frac{1}{4} \left(1 + \frac{M}{\sigma} \right)^{2}
\frac{{\cal A}}{(x^{2} - 1)^{1 + q}} 
{\left[\frac{x^{2} - 1}{x^{2} - y^{2}} \right]}^{(1 + q)^{2}} , \nonumber 
\end{eqnarray}

\noindent
where

\begin{eqnarray}
\label{q1}
{\cal A} & = & a_{+} a_{-} + b_{+} b_{-} , \nonumber \\ 
{\cal B} & = & a_{+}^{2} + b_{+}^{2} , \\
{\cal C} & = & \! (x + 1)^{q} \left[x (1 - y^{2})(\lambda + \eta) a_{+} 
+ y (x^{2} - 1)(1 - \lambda \eta) b_{+} \right] , \nonumber \\ 
\end{eqnarray}

\noindent
with

\begin{eqnarray}
\label{q2}
a_{\pm} & = & (x \pm 1)^{q} 
[x (1 - \lambda \eta) \pm (1 + \lambda \eta)] , \nonumber \\
b_{\pm} & = & (x \pm 1)^{q} [y (\lambda + \eta) \mp (\lambda - \eta)] , 
\end{eqnarray}

\noindent
and

\begin{eqnarray}
\label{q3}
\lambda & = & \alpha (x^{2} - 1)^{- q} (x + y)^{2 q} ,  \\
\eta & = & \alpha (x^{2} - 1)^{- q} (x - y)^{2 q} . \nonumber
\end{eqnarray}

\section{Relativistic Multipole Moments}

\noindent
There are several methods to get the spin and quadrupole moments from a given
metric \cite{Fodor,Guersel,Ryan}. In this section, we apply the FHP procedure 
to the QM metric. An excellent review of the FHP formalism was 
given by Filter \cite{Filter} in his diploma thesis. Filter also found
the $ {\cal S}_{11} $ component in the same way. The procedure to obtain 
the relativistic multipole moments is the following \cite{Fodor}

\begin{enumerate}
\item employ the inverse Ernst potential $ \xi^{-1} $,
\item set $ y = \cos{\theta} = 1 $, and $ \sigma x \rightarrow 1/z $ 
into $ \xi^{-1} $,
\item expand in Taylor series of $ z $ the inverse Ernst potential, and finally,
\item use the Fodor-Hoenselaers-Perj\'es (FHP) formulae \cite{Fodor}.
\end{enumerate}

\noindent
To get the relativistic multipole moment, we wrote a REDUCE program with 
the latter recipe.

% \newpage

\subsection{The Manko-Novikov Multipole Moments}

\noindent
Taking the first $ 10 $ even members of the $ \psi $ function 

\begin{equation}
\label{mnpsi}
\psi = \sum^{n = 10}_{n = 1} \alpha_{n} \frac{P_{n}}{R^{n + 1}} ,
\end{equation}

\noindent
and defining

\begin{equation}
\label{beta}
\beta^2 = {1 - \frac{a^2}{m^2}} ,
\end{equation}

\noindent
we derive the lowest relativistic multipole moments for the MN metric

\begin{eqnarray}
\label{MNMultipoles}
{\cal M}_0 & = & M \nonumber \\
{\cal S}_1 & = & S \nonumber \\
{\cal M}_2 & = & Q - S a \nonumber \\
{\cal S}_3 & = & 4 a Q - S a^2 \nonumber \\
{\cal M}_4 & = & S a^3 - \frac{4}{7} Q ({7} a^2 + {2} m^2) 
- k \alpha_4 \beta^4 m^4 \nonumber \\
{\cal S}_5 & = & S a^4 - \frac{2}{21} a Q (42 a^2 + 31 m^2) 
- 4 k \alpha_4 \beta^4 m^4 a \nonumber \\
{\cal M}_6 & = & - S a^5 
+ Q \left(4 a^4 + \frac{22}{7} m^2 a^2 + \frac{19}{33} m^4 
+ \frac{60}{77} k^3 \alpha_2 m \right) \nonumber \\
& + & \frac{1}{11} k \alpha_4 
\left(44 a^6 - 71 m^2 a^4 + 10 m^4 a^2 + 17 m^6 \right) - k \alpha_6 \beta^6 m^6  
\nonumber \\
{\cal S}_7 & = & - S a^6 
+ Q a \left(4 a^4 + \frac{490}{143} m^2 a^2 + \frac{584}{429} m^4 
- \frac{128}{143} k^3 \alpha_2 m \right) \\ 
& + & \frac{1}{143} k \alpha_4 a 
\left(572 a^6 - 482 m^2 a^4 - 752 m^4 a^2 + 662 m^6 \right) 
- 4 k \alpha_6 \beta^6 m^6 a \nonumber \\
{\cal M}_8 & = & S a^7 - Q \left(4 a^6 + \frac{1486}{429} m^2 a^4  
+ \frac{664}{429} m^4 a^2 + \frac{34}{143} m^6 \right. \nonumber \\
& - & \left. k^3 \alpha_2 \left(\frac{2368}{429} m a^2 
- \frac{4364}{3003} m^3 - \frac{40}{143} k^3 \alpha_2 \right) \right) 
+ k \alpha_4 \left(- 4 a^8 + \frac{458}{143} m^2 a^6 \right. \nonumber \\  
& + & \left. \frac{48}{11} m^4 a^4 - \frac{334}{143} m^6 a^2 - \frac{16}{13} m^8 
- \frac{226}{143} k^3 \alpha_2 \beta^4 m^5 \right) \nonumber \\   
& + & 2 k \alpha_6 \left(m^8 + 5 m^2 a^6 - 3 m^4 a^4 - m^6 a^2 - 2 a^8 \right) 
- k \alpha_8 \beta^8 m^8 \nonumber \\
{\cal S}_9 & = & S a^8 - a Q \left(4 a^6 + \frac{8586}{2431} m^2 a^4 
+ \frac{4384}{2431} m^4 a^2 + \frac{1310}{2431} m^6 \right. \nonumber \\
& - & \left. k^3 \alpha_2 
\left(\frac{21806}{2431} m a^2 + \frac{15870}{17017} m^3
+ \frac{51464}{7293} k^3 \alpha_2 \right) \right) \nonumber \\  
& - & k \alpha_4 a \left(4 a^8 - \frac{6762}{2431} m^2 a^6  
- \frac{7662}{2431} m^4 a^4 - \frac{3286}{2431} m^6 a^2 \right. \nonumber \\
& + & \left. \frac{726}{221} m^8 + \frac{5010}{2431} \beta^4 Q m^5 \right) 
\nonumber \\
& - & \frac{1}{17} k \alpha_6 a \left(68 a^8 - {94} m^2 a^6 - {126} m^4 a^4 
+ {262} m^6 a^2 - {110} m^8 \right) 
- 4 k \alpha_8 \beta^8 m^8 a \nonumber
\end{eqnarray}

\begin{eqnarray}
\label{MNMultipoles2}
{\cal M}_{10} & = & - S a^{9} 
+ Q \left(4 a^8 + \frac{9550}{2717} m^2 a^6 + \frac{5032}{2717} m^4 a^4 
+ \frac{30650}{46189} m^6 a^2  + \frac{371}{4199} m^8 \right. \nonumber \\
& - & \left. k^3 \alpha_2 \left(\frac{606490}{46189} m a^4 
+ \frac{6399608}{969969} m^3 a^2 - \frac{182530}{138567} m^5 
\right. \right. \nonumber \\
& - & \left. \left. k^3 \alpha_2 \left(\frac{459700}{323323} m^2 
- \frac{3246296}{138567} a^2 \right) \right) \right) \nonumber \\
& + & \frac{1}{46189} k \alpha_4 \left(184756 a^{10} - 131702 m^2 a^8 
- 122630 m^4 a^6 - 64624 m^6 a^4 \right. \nonumber \\
& + & \left. 100166 m^8 a^2 + 34034 m^{10} 
+ Q \left(564540 m a^6 - 1296110 m^3 a^4 + 898600 m^5 a^2 
\right. \right. \nonumber \\
& - & \left. \left. 167030 m^7 
- k^3 \alpha_2 \left(39150 a^4 - 78300 m^2 a^2 + 39150 m^4 \right) \right) 
\right) \nonumber \\ 
& + & \frac{1}{46189} \alpha_4^2 M \left(30870 a^{10} - 154350 m^2 a^8  
+ 308700 m^4 a^6 - 308700 m^6 a^4 \right. \nonumber \\ 
& + & \left. 154350 m^8 a^2 - 30870 m^{10} \right) 
+ \frac{1}{323} k \alpha_6 \left(1292 a^{10} - 1714 m^2 a^8 - 1924 m^4 a^6 
\right. \nonumber \\ 
& + & \left. 3136 m^6 a^4 - 104 m^8 a^2 - 686 m^{10} 
+ Q m (566 m^6 + 1698 m^2 a^4 - 1698 m^4 a^2 - 566 a^6) \right) \nonumber \\  
& + & \frac{1}{19} k \alpha_8 \left(76 a^{10} - 257 m^2 a^8 + 268 m^4 a^6 
- 22 m^6 a^4 - 112 m^8 a^2 + 47 m^{10} \right) \nonumber \\ 
& - & k \alpha_{10} \beta^{10} m^{10} . \nonumber
\end{eqnarray}

% \newpage

\subsection{The Quevedo-Mashhoon Multipole Moments}

\noindent
In this case, the Ernst function take the following simple form

\begin{equation}
\label{qm8}
{\xi} = \frac{(\sigma x + i a) \tanh{\psi} + m}
{(\sigma x - i a) + m \tanh{\psi}} .
\end{equation}

\noindent
Taking the first $ 10 $ even members of the function $ \psi $,

\begin{equation}
\label{qm9}
\psi = \sum^{n = 10}_{n = 1} q_{n} P_{n} Q_{n} \qquad (P_{n}(1) = 1) ,
\end{equation}

\noindent
From the ER metric, the static massive quadrupole is given by 
$ Q = {2} q M m^2 / {15} $. From the Kerr metric, one infers that the 
spin-dipole $ S = M a c $. Using these relations, and $ \beta $ as 
in (\ref{beta}), one can write the relativistic multipole moments as follows

\begin{eqnarray}
\label{qm10}
{\cal M}_0 & = & M \nonumber \\
{\cal S}_1 & = & S \nonumber \\
{\cal M}_2 & = & - S a + \beta^3 Q \nonumber \\ 
{\cal S}_3 & = & - S a^2 + 2 a \beta^3 Q \nonumber \\ 
{\cal M}_4 & = & S a^3  \left(1 + \frac{8}{21} \beta q_{2} \right)
- \frac{2}{7} \beta Q (m^2 + 9 a^2) 
+ \frac{8}{315} \beta^5 q_{4} M m^4 \nonumber \\
{\cal S}_5 & = & S a^4 \left(1 + \frac{52}{105} \beta q_{2} \right) 
- \frac{2}{7} \beta Q a (2 m^2 + 11 a^2)
+ \frac{16}{315} \beta^5 q_{4} S m^4 \nonumber \\
{\cal M}_6 & = & - S a^5 \left(1 + \frac{62}{105} \beta q_{2} 
- \frac{16}{1155} q_{2}^2 \right) \\
& + & \frac{1}{231} Q \left(24 q_{2} m^2 (3 \beta^2 a^2 - m^2) 
+ \beta (773 a^4 + 258 m^2 a^2- 8 m^4) \right) \nonumber \\
& - & \frac{8}{3465} \beta q_{4} M \left((37 a^6 + 2 m^6) 
+ 3 m^2 a^2 (11 m^2 - 24 a^2) \right) \nonumber \\
& + & \frac{16}{3003} \beta q_{6} M \left(m^6 - a^6 
- 3 \beta^2 m^4 a^2 \right) \nonumber \\
{\cal S}_7 & = & - S a^6 \left(1 + \frac{24}{35} \beta q_{2} 
- \frac{712}{19305} q_{2}^2 \right) \nonumber \\
& + & \frac{4}{9009} Q a \left(623 q_{2} m^2 (3 \beta^2 a^2 - m^2)
+ 39 \beta \left(205 a^4 + 96 m^2 a^2 - 4 m^4 \right) \right) \nonumber \\
& - & \frac{32}{3465} \beta q_{4} S \left((13 a^6 + m^6) 
+ m^2 a^2 (11 m^2 - 25 a^2) \right) \nonumber \\
& + & \frac{32}{3003} \beta q_{6} S \left(m^6 - a^6 - 3 \beta^2 m^4 a^2 \right) 
\nonumber
\end{eqnarray}

\begin{eqnarray*}
\label{qm11}
{\cal M}_8 & = & S a^7 \left(1 + \frac{2656}{3465} \beta q_{2}
- \frac{17632}{225225} q_{2}^2 - \frac{64}{96525} \beta q_{2}^3 \right. \\
& + & \left. \frac{7088}{45045} \beta q_{4} - \frac{3616}{675675} q_{2} q_{4} 
+ \frac{928}{45045} \beta q_{6} + \frac{128}{109395} \beta q_{8} \right) \\
& + & \frac{4}{3003} \beta Q \left(- 2605 a^6 - 1827 m^2 a^4 + 120 m^4 a^2 
- 4 m^6 \right) \\
& + & \frac{16}{45045} q_{2} Q \left(4922 a^6 - 4848 m^2 a^4 + 1542 m^4 a^2 
+ 37 m^6 \right) \\ 
& + & \frac{32}{6435} \beta q_{2}^2 Q \left(4 a^6 
- 6 m^2 a^4 + 4 m^4 a^2 - m^6 \right) \\
& + & \frac{16}{45045} q_{4} Q \left(452 a^6 - 678 m^2 a^4 + 452 m^4 a^2 
- 113 m^6 \right) \\
& + & \frac{16}{45045} \beta q_{4} M m^2 \left(- 830 a^6 + 329 m^2 a^4 
+ 60 m^4 a^2 - 2 m^6 \right) \\
& + & \frac{32}{45045} \beta q_{6} M m^2 \left(- 86 a^6 + 84 m^2 a^4 
- 26 m^4 a^2 - m^6 \right) \\
& + & \frac{128}{109395} \beta q_{8} M m^2 \left(- 4 a^6 + 6 m^2 a^4 - 4 m^4 a^2 
+ m^6 \right) \nonumber \\
{\cal S}_{9} & = & S a^8 \left(1 
+ \frac{2936}{3465} \beta q_{2} - \frac{171568}{1276275} q_{2}^2 
- \frac{5504}{4922775} \beta q_{2}^3 \right. \\
& - & \left. \frac{5856}{425425} q_{2} q_{4} + \frac{8768}{45045} \beta q_{4} 
+ \frac{1376}{45045} \beta q_{6} + \frac{256}{109395} \beta q_{8} \right) \\
& + & \frac{4}{3003} 
\beta Q a \left(- 2545 a^6 - 2406 m^2 a^4 + 188 m^4 a^2 - 8 m^6 \right) \\
& + & \frac{8}{85085} Q q_{2} a \left(31842 a^6 - 31188 m^2 a^4 + 9742 m^4 a^2 
+ 327 m^6 \right) \\
& + & \frac{16}{328185} \beta q_{2}^2 Q a \left(688 a^6 - 1032 m^2 a^4 
+ 688 m^4 a^2 - 172 m^6 \right) \\
& + & \frac{16}{85085} q_{4} Q a \left(2196 a^6 - 3294 m^2 a^4 + 2196 m^4 a^2 
- 549 m^6 \right) \\
& + & \frac{32}{45045} \beta q_{4} S m^2 \left(- 505 a^6 + 186 m^2 a^4 
+ 47 m^4 a^2 - 2 m^6 \right) \\
& + & \frac{32}{45045} \beta q_{6} S m^2 \left(- 127 a^6 + 123 m^2 a^4 
- 37 m^4 a^2 - 2 m^6 \right) \\
& + & \frac{256}{109395} \beta q_{8} S m^2 \left(- 4 a^6 + 6 m^2 a^4 
- 4 m^4 a^2 + m^6 \right) 
\end{eqnarray*}

\begin{eqnarray*}
\label{qm12}
{\cal M}_{10} & = & - S a^{9} \left(1 + \frac{41318}{45045} \beta q_{2} 
- \frac{78304}{373065} q_{2}^2 - \frac{65216}{26189163} \beta q_{2}^3 \right. \\
& - & \left. \frac{51552}{1616615} q_{2} q_{4} 
- \frac{1856}{4849845} \beta q_{2}^2 q_{4} 
- \frac{18112}{14549535} q_{2} q_{6} + \frac{10448}{45045} \beta q_{4} \right. \\
& - & \left. \frac{896}{2078505} q_{4}^2 + \frac{33472}{765765} \beta q_{6} 
+ \frac{10624}{2078505} \beta q_{8} + \frac{256}{969969} \beta q_{10} \right) \\
& + & \frac{1}{969969} \beta Q \left(2995461 a^8 + 4048532 m^2 a^6 
- 397840 m^4 a^4 + 27600 m^6 a^2 - 896 m^{8} \right) \\
& + & \frac{16}{2909907} q_{2} Q \left(- 843844 a^8 + 812776 m^2 a^6 
- 238254 m^4 a^4 - 17939 m^6 a^2 + 962 m^{8} \right) \\
& + & \frac{32}{8729721} \beta q_{2}^2 Q \left(- 17008 a^8 + 17082 m^2 a^6 
- 148 m^4 a^4 - 8393 m^6 a^2 + 3372 m^{8} \right) \\
& + & \frac{16}{2909907} q_{4} Q \left(- 173614 a^8 + 259486 m^2 a^6 
- 171744 m^4 a^4 + 42001 m^6 a^2 + 374 m^{8} \right) \\
& + & \frac{928}{323323} \beta q_{2} q_{4} Q \left(- 5 a^8 + 10 m^2 a^6 
- 10 m^4 a^4 + 5 m^6 a^2 - m^{8} \right) \\
& + & \frac{9056}{969969} q_{6} Q \left(- 5 a^8 + 10 a^6 m^2 - 10 a^4 m^4 
+ 5 a^2 m^6 - m^{8} \right) \\
& + & \frac{16}{14549535} \beta q_{4} M m^2 \left(378672 a^8 - 121361 m^2 a^6 
- 49730 a^4 m^4 + 3450 m^6 a^2 - 112 m^{8} \right) \\ 
& + & \frac{896}{2078505} q_{4}^2 M m^2 \left(- 5 a^8 + 10 m^2 a^6 - 10 m^4 a^4 
+ 5 m^6 a^2 - m^{8} \right) \\ 
& + & \frac{32}{14549535} \beta q_{6} M m^2 \left(58065 a^8 - 54895 m^2 a^6 
+ 15035 m^4 a^4 + 1725 a^2 m^6 - 56 m^{8} \right) \\
& + & \frac{128}{2078505} \beta q_{8} M m^2 \left(330 a^8 - 490 m^2 a^6 
+ 320 m^4 a^4 - 75 m^6 a^2 - 2 m^{8} \right) \\
& + & \frac{256}{969969} \beta q_{10} M m^2 \left(5 a^8 - 10 a^6 m^2 + 10 a^4 m^4 
- 5 a^2 m^6 + m^{8} \right) .
\end{eqnarray*}

\subsection{The $ q $-Metric Multipole Moments}

\noindent
For this metric, the relativistic multipole moments are

\begin{eqnarray}
\label{qq0}
{\cal M}_0 & = & m + q \sigma \nonumber \\
{\cal S}_1 & = & a (m + 2 q \sigma) \nonumber \\
{\cal M}_2 & = & - m^3 - 3 m^2 q \sigma - m q^2 \sigma^2 + m \sigma^2 
- \frac{1}{3} q^3 \sigma^3 + \frac{7}{3} q \sigma^3 \nonumber \\
{\cal S}_3 & = & a (3 a^2 m q^2 - a^2 m + \frac{2}{3} a^2 q^3 \sigma 
- \frac{8}{3} a^2 q \sigma - 3 m^3 q^2 - \frac{2}{3} m^2 q^3 \sigma 
- \frac{4}{3} m^2 q \sigma) \nonumber \\
{\cal M}_4 & = & m^5 + 5 m^4 q \sigma + 6 m^3 q^2 \sigma^2 - 2 m^3 \sigma^2 
+ \frac{46}{21} m^2 q^3 \sigma^3 - \frac{166}{21} m^2 q \sigma^3 
+ \frac{19}{21} m q^4 \sigma^4 \nonumber \\
& - & \frac{16}{3} m q^2 \sigma^4 + m \sigma^4 
+ \frac{19}{105} q^5 \sigma^5 - \frac{16}{21} q^3 \sigma^5 
+ \frac{43}{15} q \sigma^5 \nonumber \\
{\cal S}_5 & = & a \left(\frac{145}{63} a^4 m q^4 - \frac{520}{63} a^4 m q^2 
+ a^4 m + \frac{74}{315} a^4 q^5 \sigma - \frac{52}{63} a^4 q^3 \sigma 
+ \frac{46}{15} a^4 q \sigma \right. \nonumber \\
& - & \left. \frac{290}{63} a^2 m^3 q^4 
+ \frac{410}{63} a^2 m^3 q^2 - \frac{148}{315} a^2 m^2 q^5 \sigma 
- \frac{256}{63} a^2 m^2 q^3 \sigma + \frac{316}{105} a^2 m^2 q \sigma \right. 
\nonumber \\ 
& + & \left. \frac{145}{63} m^5 q^4 + \frac{110}{63} m^5 q^2 
+ \frac{74}{315} m^4 q^5 \sigma + \frac{44}{9} m^4 q^3 \sigma 
- \frac{8}{105} m^4 q \sigma \right) \\
{\cal M}_6 & = & - m^7 - 7 m^6 q \sigma - 15 m^5 q^2 \sigma^2 + 3 m^5 \sigma^2 
- \frac{775}{63} m^4 q^3 \sigma^3 + \frac{1093}{63} m^4 q \sigma^3 \nonumber \\ 
& - & \frac{125}{21} m^3 q^4 \sigma^4 + \frac{556}{21} m^3 q^2 \sigma^4 
- 3 m^3 \sigma^4 - \frac{11}{5} m^2 q^5 \sigma^5 
+ \frac{8524}{693} m^2 q^3 \sigma^5 \nonumber \\ 
& - & \frac{46997}{3465} m^2 q \sigma^5 
- \frac{389}{495} m q^6 \sigma^6 + \frac{127}{33} m q^4 \sigma^6 
- \frac{39397}{3465} m q^2 \sigma^6 + m \sigma^6 
- \frac{389}{3465} q^7 \sigma^7 \nonumber \\
& + & \frac{719}{3465} q^5 \sigma^7 - \frac{37}{55} q^3 \sigma^7 
+ \frac{337}{105} q \sigma^7 \nonumber \\
{\cal S}_7 & = & a \left(\frac{27073}{19305} a^6 m q^6 
- \frac{21427}{3861} a^6 m q^4 + \frac{287647}{19305} a^6 m q^2 - a^6 m 
+ \frac{2986}{45045} a^6 q^7 \sigma \right. \nonumber \\ 
& - & \left. \frac{316}{6435} a^6 q^5 \sigma + \frac{94}{195} a^6 q^3 \sigma 
- \frac{352}{105} a^6 q \sigma - \frac{27073}{6435} a^4 m^3 q^6 
+ \frac{1340}{429} a^4 m^3 q^4 \right. \nonumber \\
& - & \left. \frac{54947}{6435} a^4 m^3 q^2 
- \frac{2986}{15015} a^4 m^2 q^7 \sigma - \frac{7496}{1287} a^4 m^2 q^5 \sigma 
+ \frac{127454}{6435} a^4 m^2 q^3 \sigma \right. \nonumber \\
& - & \left. \frac{3368}{693} a^4 m^2 q \sigma + \frac{27073}{6435} a^2 m^5 q^6 
+ \frac{1217}{117} a^2 m^5 q^4 - \frac{14206}{2145} a^2 m^5 q^2 
\right. \nonumber \\ 
& + & \left. \frac{2986}{15015} a^2 m^4 q^7 \sigma 
+ \frac{75908}{6435} a^2 m^4 q^5 \sigma - \frac{116924}{6435} a^2 m^4 q^3 \sigma 
+ \frac{256}{1155} a^2 m^4 q \sigma \right. \nonumber \\
& - & \left. \frac{27073}{19305} m^7 q^6 
- \frac{30794}{3861} m^7 q^4 + \frac{5048}{19305} m^7 q^2 
- \frac{2986}{45045} m^6 q^7 \sigma 
- \frac{12704}{2145} m^6 q^5 \sigma \right. \nonumber \\ 
& - & \left. \frac{4544}{2145} m^6 q^3 \sigma 
- \frac{32}{3465} m^6 q \sigma \right) \nonumber
\end{eqnarray}

\begin{eqnarray}
\label{qq1}
{\cal M}_8 & = & m^9 + 9 m^8 q \sigma + 28 m^7 q^2 \sigma^2 - 4 m^7 \sigma^2 
+ \frac{3812}{99} m^6 q^3 \sigma^3 - \frac{3044}{99} m^6 q \sigma^3 \nonumber \\ 
& + & \frac{938}{33} m^5 q^4 \sigma^4 - \frac{2480}{33} m^5 q^2 \sigma^4 
+ 6 m^5 \sigma^4 + \frac{32938}{2145} m^4 q^5 \sigma^5 
- \frac{30896}{429} m^4 q^3 \sigma^5 \nonumber \\
& + & \frac{27154}{715} m^4 q \sigma^5 
+ \frac{821452}{135135} m^3 q^6 \sigma^6 - \frac{932884}{27027} m^3 q^4 \sigma^6 
+ \frac{8883508}{135135} m^3 q^2 \sigma^6 \nonumber \\
& - & 4 m^3 \sigma^6 + \frac{24844}{10395} m^2 q^7 \sigma^7 
- \frac{1630604}{135135} m^2 q^5 \sigma^7 
+ \frac{4516228}{135135} m^2 q^3 \sigma^7 \nonumber \\ 
& - & \frac{887132}{45045} m^2 q \sigma^7 
+ \frac{257}{385} m q^8 \sigma^8 - \frac{96568}{45045} m q^6 \sigma^8 
+ \frac{100322}{15015} m q^4 \sigma^8 \nonumber \\
& - & \frac{836312}{45045} m q^2 \sigma^8 + m \sigma^8 
+ \frac{257}{3465} q^9 \sigma^9 - \frac{2152}{45045} q^7 \sigma^9 
- \frac{12874}{45045} q^5 \sigma^9 \nonumber \\
& - & \frac{5128}{45045} q^3 \sigma^9 
+ \frac{1091}{315} q \sigma^9 \nonumber \\
{\cal S}_9 & = & a \left(\frac{177511}{255255} a^8 m q^8 
- \frac{26024}{12155} a^8 m q^6 + \frac{1912994}{255255} a^8 m q^4 
- \frac{1918792}{85085} a^8 m q^2 + a^8 m \right. \nonumber \\
& - & \left. \frac{20486}{6891885} a^8 q^9 \sigma 
+ \frac{59144}{328185} a^8 q^7 \sigma 
- \frac{149956}{208845} a^8 q^5 \sigma + \frac{155192}{530145} a^8 q^3 \sigma 
+ \frac{1126}{315} a^8 q \sigma \right. \nonumber \\
& - & \left. \frac{710044}{255255} a^6 m^3 q^8 
- \frac{1751252}{255255} a^6 m^3 q^6 + \frac{2530868}{85085} a^6 m^3 q^4 
+ \frac{97892}{12155} a^6 m^3 q^2 \right. \nonumber \\
& + & \left. \frac{81944}{6891885} a^6 m^2 q^9 \sigma 
- \frac{13208144}{2297295} a^6 m^2 q^7 \sigma 
+ \frac{53663312}{2297295} a^6 m^2 q^5 \sigma \right. \nonumber \\ 
& - & \left. \frac{346745888}{6891885} a^6 m^2 q^3 \sigma 
+ \frac{61448}{9009} a^6 m^2 q \sigma + \frac{355022}{85085} a^4 m^5 q^8 
+ \frac{568852}{17017} a^4 m^5 q^6 \right. \nonumber \\
& - & \left. \frac{6788302}{85085} a^4 m^5 q^4 
+ \frac{265388}{17017} a^4 m^5 q^2 - \frac{40972}{2297295} a^4 m^4 q^9 \sigma 
+ \frac{12380128}{765765} a^4 m^4 q^7 \sigma \right. \nonumber \\
& - & \left. \frac{4893500}{153153} a^4 m^4 q^5 \sigma 
+ \frac{8150456}{208845} a^4 m^4 q^3 \sigma 
- \frac{6416}{15015} a^4 m^4 q \sigma - \frac{710044}{255255} a^2 m^7 q^8 
\right. \nonumber \\
& - & \left. \frac{3208596}{85085} a^2 m^7 q^6 
+ \frac{1471432}{36465} a^2 m^7 q^4 - \frac{8864}{7735} a^2 m^7 q^2 
+ \frac{81944}{6891885} a^2 m^6 q^9 \sigma \right. \nonumber \\
& - & \left. \frac{36312368}{2297295} a^2 m^6 q^7 \sigma 
- \frac{7586872}{2297295} a^2 m^6 q^5 \sigma 
+ \frac{79889104}{6891885} a^2 m^6 q^3 \sigma 
+ \frac{1504}{45045} a^2 m^6 q \sigma \right. \nonumber \\
& + & \left. \frac{177511}{255255} m^9 q^8 
+ \frac{260828}{19635} m^9 q^6 + \frac{16948}{7735} m^9 q^4 
+ \frac{4112}{85085} m^9 q^2 - \frac{20486}{6891885} m^8 q^9 \sigma 
\right. \nonumber \\ 
& + & \left. \frac{2393224}{459459} m^8 q^7 \sigma 
+ \frac{4139368}{328185} m^8 q^5 \sigma - \frac{825152}{1378377} m^8 q^3 \sigma 
- \frac{64}{45045} m^8 q\sigma \right) \nonumber 
\end{eqnarray}

\begin{eqnarray}
\label{qq3}
{\cal M}_{10} & = & - m^{11} - 11 m^{10} q \sigma - 45 m^9 q^2 \sigma^2 
+ 5 m^9 \sigma^2 - \frac{38075}{429} m^8 q^3 \sigma^3 
+ \frac{20645}{429} m^8 q \sigma^3 \nonumber \\
& - & \frac{41510}{429} m^7 q^4 \sigma^4 
+ \frac{70040}{429} m^7 q^2 \sigma^4 - 10 m^7 \sigma^4 
- \frac{150698}{2145} m^6 q^5 \sigma^5 + \frac{35528}{143} m^6 q^3 \sigma^5 
\nonumber \\ 
& - & \frac{176062}{2145} m^6 q \sigma^5 - \frac{27070}{693} m^5 q^6 \sigma^6 
+ \frac{1752710}{9009} m^5 q^4 \sigma^6 
- \frac{1966522}{9009} m^5 q^2 \sigma^6 \nonumber \\ 
& + & 10 m^5 \sigma^6 
- \frac{884582}{51051} m^4 q^7 \sigma^7 
+ \frac{25565266}{255255} m^4 q^5 \sigma^7 
- \frac{11654186}{51051} m^4 q^3 \sigma^7 \nonumber \\ 
& + & \frac{17219134}{255255} m^4 q \sigma^7 
- \frac{3200647}{459459} m^3 q^8 \sigma^8 
+ \frac{16620592}{459459} m^3 q^6 \sigma^8 
- \frac{48051098}{459459} m^3 q^4 \sigma^8 \nonumber \\  
& + & \frac{58197248}{459459} m^3 q^2 \sigma^8 
- 5 m^3 \sigma^8 - \frac{66272933}{26189163} m^2 q^9 \sigma^9 
+ \frac{12124624}{1247103} m^2 q^7 \sigma^9 \nonumber \\  
& - & \frac{139964914}{4849845} m^2 q^5 \sigma^9 
+ \frac{14896480}{220077} m^2 q^3 \sigma^9 
- \frac{380725673}{14549535} m^2 q \sigma^9 
- \frac{443699}{793611} m q^{10} \sigma^{10} \nonumber \\ 
& + & \frac{83879}{73359} m q^8 \sigma^{10} 
- \frac{49852798}{43648605} m q^6 \sigma^{10} 
+ \frac{62188166}{8729721} m q^4 \sigma^{10} 
- \frac{387595759}{14549535} m q^2 \sigma^{10} \nonumber \\ 
& + & m \sigma^{10} - \frac{443699}{8729721} q^{11} \sigma^{11} 
+ \frac{522479}{26189163} q^9 \sigma^{11} 
+ \frac{19539686}{43648605} q^7 \sigma^{11} 
- \frac{3438746}{2567565} q^5 \sigma^{11} \nonumber \\ 
& + & \frac{110793707}{130945815} q^3 \sigma^{11} 
+ \frac{12701}{3465} q \sigma^{11} \nonumber
\end{eqnarray}

\noindent
The MN, the QM metrics might be useful to represent the spacetime of real 
non-collapsed body due to their large freedom in choosing the multipole 
moments of the exterior gravitational field. Nevertheless, the higher order 
spin moments cannot be chosen freely but are totally determined by the set of 
metric parameters. The $ q $-metric may be useful to represent the spacetime 
of deformed objects, but again the higher order spin moments cannot be chosen 
freely.

\noindent
In next section, we compare the static multipoles of the MN metric derived 
from (\ref{MNMultipoles}) and the ER multipole moments deduced from 
(\ref{qm10}) with those ones of different static metrics.

%\newpage

\section{Comparisons with Static Metrics}

\noindent
For a comparison with other static metrics, we consider the 
Gut\-su\-nayev-Manko (GM) \cite{GM}, the Hern\'andez-Mart{\'{\i}}n 
spacetimes (HM I and II) \cite{HM}, and the static $ q $-metric 
\cite{Quevedo5,FQP}. These metrics are static solutions of the EFE and 
might be related with the external gravitational field of a body with mass 
and quadrupole moment. The differences of the metrics are in the fields 
$ \psi $ and $ \gamma $. To obtain the multipole moments of these metrics, 
the FHP procedure was also employed. In the case of the generalized ER metric, 
we have from (\ref{qm10}) the following mass multipoles

\begin{eqnarray*}
\label{ERRMM}
{\cal M}_0 & = & M \\
{\cal M}_2 & = & Q \\
{\cal M}_4 & = & \frac{2}{21} \left(- 3 Q + 60 q_{4} M m^2 \right) m^2 \\
{\cal M}_6 & = & - \frac{8}{3003} \left(13 (1 + 3 q_{2}) Q 
+ 30 (13 q_{4} - 15 q_{6}) M m^2 \right) m^4 \\
{\cal M}_8 & = & - \frac{16}{765765} \left((255 - 629 q_{2} + 238 q_{2}^2 
+ 1921 q_{4}) Q \right. \\
& + & \left. 30 (255 q_{4} + 255 q_{6} - 420 q_{8}) M m^2 \right) m^6 \\
{\cal M}_{10} & = & - \frac{32}{2909907} \left(
(84 - 481 q_{2} - 1124 q_{2}^2 - 187 q_{4} + 261 q_{2} q_{4} + 849 q_{6}) Q 
\right. \\
& + & \left. 30 (84 q_{4} + 294 q_{4}^2 + 84 q_{6} + 84 q_{8} - 180 q_{10}) M m^2 
\right) m^8 .
\end{eqnarray*}

\noindent
In the case of the MN metric if $ \alpha = a = 0 $ , we have

\begin{eqnarray}
\label{MNMultipoles3}
{\cal M}_0 & = & M = k \nonumber \\
{\cal M}_2 & = & Q \nonumber \\
{\cal M}_4 & = & - \frac{8}{7} Q m^2 - \alpha_4 M m^4 \nonumber \\
{\cal M}_6 & = & Q m \left(\frac{19}{33} m^3 
+ \frac{60}{77} k^3 \alpha_2 \right) 
+ \frac{1}{11} M m^6 ({17} \alpha_4 - 11 \alpha_6) \nonumber \\
{\cal M}_8 & = & - Q \left(\frac{34}{143} m^6 
+ k^3 \alpha_2 \left(\frac{4364}{3003} m^3 
+ \frac{40}{143} k^3 \alpha_2 \right) \right) \\
& - & \alpha_4 M m^5 \left(\frac{16}{13} m^3 
+ \frac{226}{143} k^3 \alpha_2 \right) 
+ M m^8 (2 \alpha_6 - \alpha_8) \nonumber \\
{\cal M}_{10} & = & Q m^2 \left(\frac{371}{4199} m^6 
+ k^3 \alpha_2 \left(\frac{182530}{138567} m^3 
+ k^3 \alpha_2 \frac{459700}{323323} \right) \right) \nonumber \\
& + & \frac{1}{46189} \alpha_4 M m^4 \left(34034 m^{6} 
+ k^3 \alpha_2 \left(167030 m^3 
+ 39150 k^3 \alpha_2 \right) \right) \nonumber \\ 
& - & \frac{30870}{46189} \alpha_4^2 M m^{10} 
- \frac{1}{323} \alpha_6 M m^7 (686 m^{3} + 566 k^3 \alpha_2) \nonumber \\  
& + & \frac{1}{19} M m^{10} ({47} \alpha_8 - 19 \alpha_{10}) . \nonumber
\end{eqnarray}

\noindent
For the GM metric \cite{GM} and the HM \cite{HM}, one uses the 
following Ernst potential

\begin{eqnarray}
\label{EGMHM}
{\xi} & = & \frac{x \tanh{\psi} + 1}{x + \tanh{\psi}} .
\end{eqnarray}

\noindent
The field $ \psi $ for the GM is given by

\begin{eqnarray}
\label{psigamgm}
\psi & = & {q} \frac{x}{(x^2 - y^2)^3} (x^2 - 3 x^2 y^2 + 3 y^2 - y^4) 
\end{eqnarray}

\noindent
and the lowest six mass multipoles read

\begin{eqnarray*}
\label{GMRMM}
{\cal M}_0 & = & M \\
{\cal M}_2 & = & Q \\
{\cal M}_4 & = & \frac{6}{7} Q m^2 \\
{\cal M}_6 & = & \frac{8}{231} Q m^4 \left(14 - 45 q \right) \\
{\cal M}_8 & = & \frac{8}{3003} Q m^6 \left(84 - 1282 q - 420 q^2 \right) \\
{\cal M}_{10} & = & \frac{32}{3927} Q m^8 \left(\frac{2772}{247} 
- \frac{1343804}{2717} q - \frac{2550}{19} q^2 \right) ,
\end{eqnarray*}

\noindent
where $ Q = 2 q M m^2 $.

\noindent
For static metrics, Hern\'andez and Mart{\'{\i}}n \cite{HM} have shown 
that the Weyl moments $ a_{l} $, defined by

\begin{eqnarray*}
\label{Weyl}
\psi & = & \sum^{\infty}_{l = 0} \frac{a_{l}}{r^{l + 1}} P_{l}(\cos{\theta}) 
\end{eqnarray*}

\noindent
($ (r, \, \theta) $ are spherical Weyl coordinates), can be found if the 
relativistic multipole moments are given and vice versa. It is easy to see that 
for a stationary metric, one can invert the FHP formulas to find the metric 
coefficients. We have written a REDUCE program (available upon request) to 
invert the FHP formulas. Thus, the metric structure can be determined in 
principle from the relativistic multipole moments. There are two HM solutions. 
In case of the first HM metric, the field $ \psi $ is given by

\begin{eqnarray}
\label{psigamhm}
\psi & = & \frac{5}{8} {q} 
\left[\frac{1}{4} \left((3 x^2 - 1) (3 y^2 - 1) - 4 \right) 
\ln{\left[\frac{x - 1}{x + 1}\right]} - \frac{2 x}{(x^2 - y^2)} 
+ \frac{3}{2} x (3 y^2 - 1) \right]  \nonumber
\end{eqnarray}

\noindent
and the lowest mass multipoles are given by

\begin{eqnarray*}
\label{HMRMM}
{\cal M}_0 & = & M \\
{\cal M}_2 & = & Q \\
{\cal M}_4 & = & 0 \\
{\cal M}_6 & = & - \frac{60}{77} q Q m^4 \\
{\cal M}_8 & = & - \frac{4}{3003} q Q m^6 \left(265 + 210 q \right) \\
{\cal M}_{10} & = & \frac{4}{3927} q Q m^8 
\left(- \frac{34790}{247} + \frac{769125}{1729} q \right) ,
\end{eqnarray*}

\noindent
where $ Q = q M m^2 $.

\noindent
For the second HM metric, the field $ \psi $ is given by (\ref{psigamhm}) plus 
a second term with the parameter $ q^2 $, see \cite{HM,FQP}. 
The lowest six mass multipoles are given by

\begin{eqnarray*}
\label{HMRMMII}
{\cal M}_0 & = & M \\
{\cal M}_2 & = & Q \\
{\cal M}_4 & = & 0 \\
{\cal M}_6 & = & 0 \\
{\cal M}_8 & = & - \frac{40}{143} q^2 Q m^6 \\
{\cal M}_{10} & = & - \frac{42140}{46189} q^2 Q m^8 .
\end{eqnarray*}

\noindent
Finally from (\ref{q0}), we get the following mass multipoles for 
the $ q $-metric 

\begin{eqnarray*}
\label{ZVRMM}
{\cal M}_0 & = &  \delta m \\
{\cal M}_2 & = & Q = \frac{1}{3} \delta m^3 (1 - \delta^2) \\
{\cal M}_4 & = & \delta m^5
\left(\frac{19}{105} \delta^4 - \frac{8}{21} \delta^2 + \frac{1}{5} \right) \\
{\cal M}_6 & = & \delta m^7 \left(
- \frac{389}{3465} \delta^6 + \frac{23}{63} \delta^4 
- \frac{457}{1155} \delta^2 + \frac{1}{7} \right) \\
{\cal M}_8 & = & \delta  m^9 \left(
\frac{257}{3465} \delta^8 - \frac{44312}{135135} \delta^6 
+ \frac{73522}{135135} \delta^4 - \frac{54248}{135135} \delta^2 + \frac{1}{9} 
\right) \\
{\cal M}_{10} & = & \delta m^{11} \left(
- \frac{443699}{8729721} \delta^{10} + \frac{17389}{61047} \delta^8 
- \frac{27905594}{43648605} \delta^6 + \frac{6270226}{8729721} \delta^4 
\right. \\
& - & \left. \frac{5876077}{14549535} \delta^2 + \frac{1}{11}\right) \ ,
\end{eqnarray*}

\noindent
where $ \delta = 1 + q $.

\noindent
The multipole structures of the ER, the static MN, the GM, the HM (I and II) 
and $ q $-metrics are different and the spacetimes are not isometric to each 
other. For modelling the exterior metric of a realistic static body 
the generalized ER or QM metric and MN metric with their infinitely many 
independent mass moments $ q_{n} $ and $ \alpha_{n} $ that can be chosen freely 
might be employed.

%\newpage

\section{Conclusions}

\noindent
In this contribution, we gave a brief review of the QM spacetime, and the MN 
metric. These metrics contain an infinite set of multipole moments 
that can be chosen freely. The rotating $ q $-metric was also introduced. 
This one has only three parameters mass, rotation parameter and quadrupole. 
Therefore, it may be useful to represent deformed objects. The relativistic 
multipole moments for these metrics were derived using the corresponding Ernst 
function and the FHP formalism. The Ernst function takes a simple form for even 
$ \alpha_{n} $ and $ q_{n} $ and then it allows us to deduce the GH moments. 
In principle, it is possible to extent the REDUCE program to include zonal 
harmonics with odd values of $ n $ (that violate equatorial symmetry). 

\noindent 
Although the MN and the QM metrics have a large set of multipole moment 
parameters for the exterior gravitational field, their higher order spin 
moments cannot be chosen freely, because they are totally determined by the set 
of metric parameters. The GH moments for the generalized ER metric and those of 
the static MN metric were compared with those of other metrics, namely the GM, 
HM (I and II), and $ q $-metrics. All compared spacetimes are not isometric 
each other. The generalized ER or QM and the MN metrics are more appropriate to 
describe the exterior gravitational field of a real static object due to their 
more general multipole structures. 

% References

\end{document}